\documentclass[a4paper,dvips 2t,superscriptaddress,showkeys]{revtex4} 
\usepackage{epsfig,graphics,graphicx,amsmath,amssymb}
\usepackage{graphicx}
\usepackage{subfigure}
\usepackage{color}
\begin{document}

\title{\bf\bf  Annihilation of singlet fermionic dark matter into two photons via pseudo-scalar mediator   }
\author{ M.M. Ettefaghi }\email{mettefaghi@qom.ac.ir}\affiliation{Department of Physics, University of Qom, Ghadir Blvd., Qom 371614-6611, I.R. Iran}
\author{ R. Moazzemi }\email{r.moazzemi@qom.ac.ir}\affiliation{Department of Physics, University of Qom, Ghadir Blvd., Qom 371614-6611, I.R. Iran}
\author{ M. Yazdani Najafabadi }\email{maryamyazdani8@yahoo.com}\affiliation{Department of Physics, University of Qom, Ghadir Blvd., Qom 371614-6611, I.R. Iran}

\begin{abstract}
	We consider the indirect detection of dark matter within an extension of the standard model (SM) including a singlet fermion as cold dark matter (CDM) and a singlet pseudo-scalar as a mediator between dark matter and the SM particles. The annihilation cross section of the CDM into two monochromatic photons is calculated and compared with the latest H.E.S.S. data. Although for dark matter masses below 1 TeV the predicted observable cross sections are far from the sensitivity of the recent gamma-ray experiments, 
	it can be comparable to the strongest H.E.S.S. upper bounds for some models with more massive CDM.
\end{abstract}

\maketitle

\section{Introduction} 
Cosmological and astrophysical observations indicate that the greatest amount of matter in the universe is unknown. All these observations are due to the collective gravitational effects and, so far, there is not any confirmed signal which determines the nature of this invisible matter. This kind of matter is so called dark matter. The Planck experiment, based on high precision measurement of the cosmic microwave background (CMB), recently obtained the following value for the dark matter relic density \cite{1}: 
\begin{equation}
\Omega_{DM}h^{2}= 0.1196\pm0.0031,\label{omega}
\end{equation}
where $h \approx 0.7$ is the scaled Hubble constant in the units of 100
km/sec/Mpc. The search for finding an appropriate explanation for dark matter problem is one of the most outstanding problems in both particle physics and the cosmology and astrophysics. Particle physicists believe that the solution to the mystery of dark matter should probably lie in the domain of elementary particles. One cannot find a proper candidate for the explanation of dark matter in the content of the standard model (SM). Therefore, dark matter is one of the most important motivations to extend the SM. A particle as a proper candidate for dark matter needs to satisfy some important constraints and properties which have been discussed in \cite{2}. In particular, 
\begin{itemize}
	\item It has to have an appropriate weak interaction with the SM particles such that its relic abundance satisfies the observed constrain (\ref{omega}). These candidates are called weakly interacting massive particles (WIMP).
	\item It has to be consistent with the direct detection bounds, e.g. \cite{lux,xenon,pandax}. These experiments have been devised to detect a dark matter particle directly but have not succeed until now. 
	\item It should meet the constraints of indirect detection. Most importantly, gamma-ray searches such as Fermi-LAT \cite{fermi} and H.E.S.S. \cite{8} have determined other bounds on dark matter candidates. 
\end{itemize}
Moreover, among the various scenarios to structure formation which have been pursued, cold dark matter (CDM) is the most successful. Indeed, CDM refers to the very slowly moving elementary particles left over from the earliest moments. There are many models, beyond the SM,  in the literature to predict a candidate for CDM. For instance, supersymmetric models with R parity \cite{3,4},  the extra dimensional models with conserved Kaluza-Klein parity \cite{5}, the T-parity  conserved little Higgs model \cite{6} and so on provide CDM candidates. These models, besides of CDM, introduce others degrees of freedom which have not been yet confirmed experimentally. Meanwhile, in order to have a massive fermion as CDM candidate, some authors have extended economically the SM by a singlet fermion to play CDM role and either a singlet scalar \cite{kim,bagherian,10} or pseudo-scalar \cite{7,Bastian,77} as mediator. In the case of singlet scalar, while almost the whole parameter space is excluded by the recent direct experimental CDM survey bound \cite{bagherian,10}, the annihilation cross section of CDM into two photons is below the recent indirect experimental data such as Fermi-LAT and H.E.S.S. \cite{11,moazzemi}. In the case of  pseudo-scalar mediator, the tree level scattering cross section of CDM from nucleons is suppressed and consequently is far below the recent direct detection bounds \cite{7}, however, it was shown that its leading loop level  contribution will be sizeable by future direct detection experiments such as XENONnT \cite{77}.

In this paper, we study the indirect survey of singlet fermionic CDM with pseudo-scalar mediator via calculating the annihilation cross section of CDM into two photons.  We respect the unitarity, relic density, vacuum stability and  perturbativity constraints and perform this for the CDM masses up to 10 TeV. Then the results are compared with the strongest upper bound reported by H.E.S.S. \cite{8}. We do not consider data reported by Fermi-LAT since they are for the CDM masses up to 200 GeV \cite{fermi}. In this region, the predicted values by our considered models are so below the experimental bounds. 

The paper is organized as follows: in the next section we give a brief review of the model. The annihilation cross section of CDM into two photons is obtained and compared with the H.E.S.S. data in the third section. We summarize our results in the last section.
\section{The model} 
One of the renormalizable extensions  of the SM Lagrangian, including a DM candidate, is achieved by adding two new extra fields: a gauge singlet fermion as CDM and a real singlet pseudo-scalar field as mediator. The fermion field interact with the SM particles only via the singlet Higgs portal. On the other hand, because the lepton and baryon numbers of this new fermion are taken to be zero,  there is not any mixing between the new fermion and the SM ones. The Lagrangian of this model is as fallows \cite{7}:
\begin{equation}
\mathcal{L}=\mathcal{L}_{\rm{SM}}+\mathcal{L}_{\rm{hid}}+\mathcal{L}_{\rm{int}},\label{tot}
\end{equation} 
where 
$\mathcal{L}_{\rm{SM}}$
is the SM Lagrangian and 
$\mathcal{L}_{\rm{hid}}$ includes the kinetic terms related to the singlet fermion and the singlet pseudo-scalar, the potential term of the singlet pseudo-scalar and the interaction between these two new particles:   
\begin{equation}
\mathcal{L}_{\rm{hid}}=\mathcal{L}_{\chi}+\mathcal{L}_{\phi}-g_s \overline{\chi}\gamma^{5}\chi \phi.
\end{equation}
Here, $\mathcal{L}_{\chi}$ and $\mathcal{L}_{\phi}$ are defined as follows:
\begin{equation}
\mathcal{L}_{\chi}=\overline{\chi}(i\!\not\!{\partial}-m_{\chi})\chi,
\end{equation}
and
\begin{equation}
\mathcal{L}_{\phi}=\frac{1}{2}(\partial_{\mu}\phi)^{2}-\frac{1}{2}{m_{0}}^{2}\phi^{2}-\frac{\lambda}{4!}\phi^{4}.
\end{equation}
In Eq. \eqref{tot}, $\mathcal{L}_{\rm{int}}$ 
is the interaction Lagrangian between the singlet pseudo-scalar and SM-Higgs doublet 
\begin{equation}
\mathcal{L}_{\rm{int}} =- \lambda_{2}H^{\dag}H \phi^{2}.
\end{equation}
 Under the parity transformation
$\phi(t,\mathbf{x }) \to -\phi(t,-\mathbf{x }) $
and
$\chi(t, \mathbf{x }) \to  \gamma^{0} \chi (t,-\mathbf{x }) $, the Lagrangian $\mathcal{L}$ is invariant and consequently  is \textit{CP}-invariant (since there are no terms such as 
$\phi$, $\phi^{3}$ , $H^{\dag}H\phi$ and so on). The vacuum expectation value (VEV) of the SM Higgs is non-zero which leads to the electroweak spontaneous symmetry breaking. This means that
\begin{equation}
	 \langle H \rangle=\frac{1}{\sqrt{2}}\begin{pmatrix}0\\
v_{H}\\
\end{pmatrix}.
\end{equation}
The singlet pseudo-scalar can also have a non-zero VEV, $\langle \phi \rangle=v_{\phi}$. If it is so, it will lead to CP violation although the Lagrangian (\ref{tot}) is written CP invariantly.
After symmetry breaking, the SM Higgs doublet and the singlet pseudo-scalar are written as follows:
\begin{equation}
H=\frac{1}{\sqrt{2}}\begin{pmatrix}0\\
\tilde{h}+v_{H}\\
\end{pmatrix},
\end{equation}
and
\begin{equation}
\phi=S+v_{\phi},
\end{equation}
where new fields $\tilde{h}$ and $S$ denote the fluctuation around the corresponding VEVs. By diagonalizing the Higgs mass matrix, one can find the following mass eigenstates fields:
\begin{equation*}
h=\sin\theta S+ \cos\theta\tilde{h}, 
\end{equation*} 
\begin{equation}
\rho=\cos\theta S-\sin\theta \tilde{h},
\end{equation} 
where $\theta$ depends on the VEVs and couplings, and is not an independent parameter \cite{7},
 $h$ and $\rho$ stand for the SM Higgs with mass $m_{h}=125 $ GeV \cite{pdg2022} and singlet pseudo-scalar Higgs with unknown mass $m_\rho$, respectively. Moreover, $v_{H}$ is known to be $246 $ GeV, while $v_{\phi}$ is unknown.

In this theory, we want the singlet fermion to have the CDM role. Therefore, its relic abundance must be consistent with the Plank observation (\ref{omega}).  The relic abundance of the singlet fermion is determined by its annihilation into  other particles in early universe. The annihilation cross section of singlet fermion into a pair of massive fermions, a pair of massive gauge bosons, and two and three Higgs bosons accommodated in this theory at the leading order is given by 
\begin{equation}
	\sigma_{\rm{ann}}(s)=\sigma_{\rm{SM}}
	+\sigma _{\rm{2 Higgs}}+\sigma _{\rm{3 Higgs}}.
	\label{1}
\end{equation}
 At the leading order, the annihilation into SM particles (except for two Higgs particles) and three Higgs particles is performed only through the $s$-channel. Meanwhile, the annihilation into two Higgs particles is performed via the $s$-, $t$- and $u$-channels. An explicit expressions for $\sigma_{\rm{SM}}$ and $\sigma _{\rm{2 Higgs}}$ is given in Appendix. We ignore annihilation into three Higgs bosons because it is suppressed in comparison to the two Higgs ones due to the smaller phase space integral and also smaller vertex couplings after symmetry breaking. The mass of the SM Higgs, $m_h$, is
$ 125.25 $  GeV and its decay width  is about
$\Gamma^{SM} _{Higgs} =3.2 $  MeV\cite{pdg2022}.

\section{The relic density}  
\begin{figure}[th]
	\centering
	\subfigure[]
	{\includegraphics[scale=0.37]{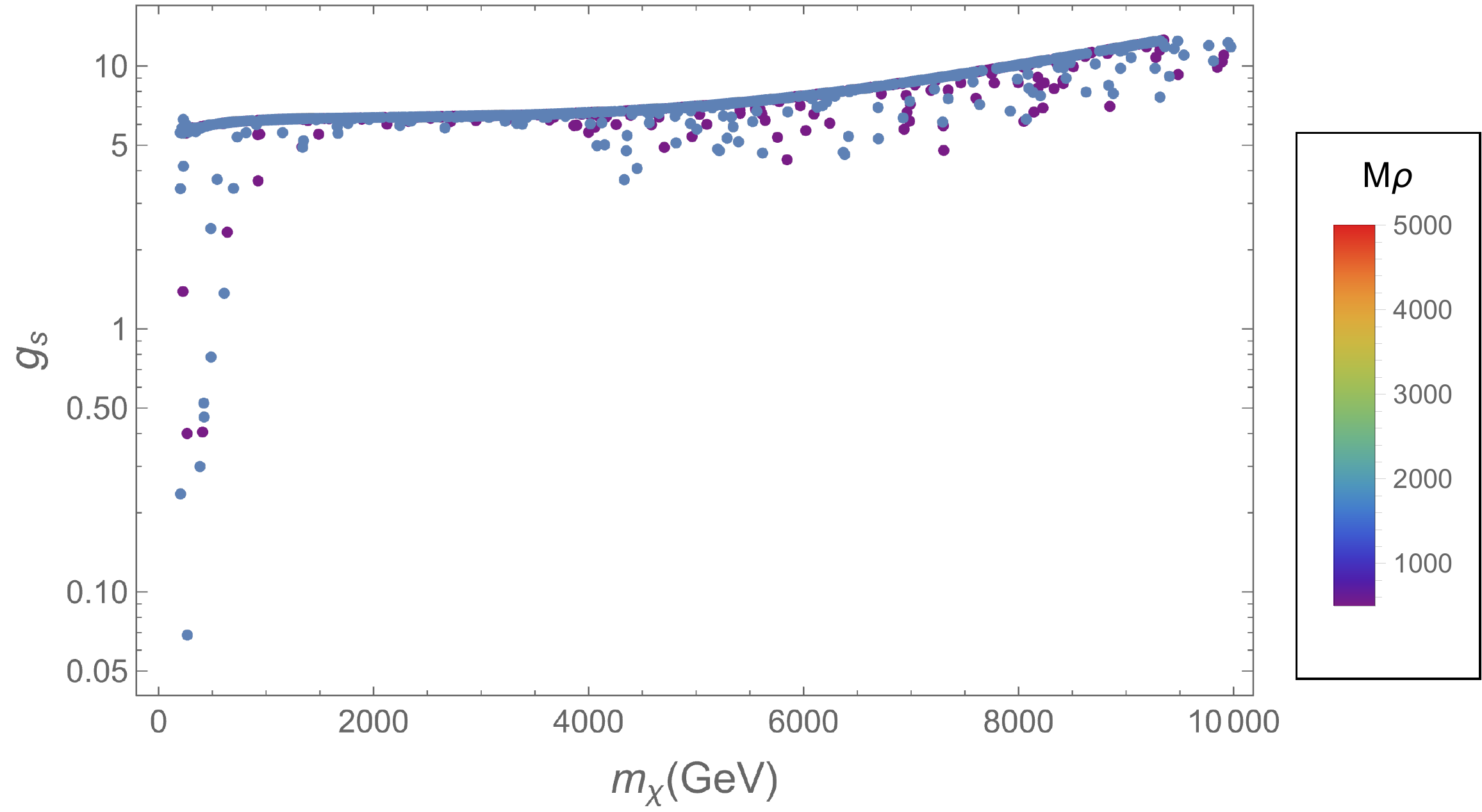}}\hspace{0.3cm}\subfigure[]{\includegraphics[scale=0.35]{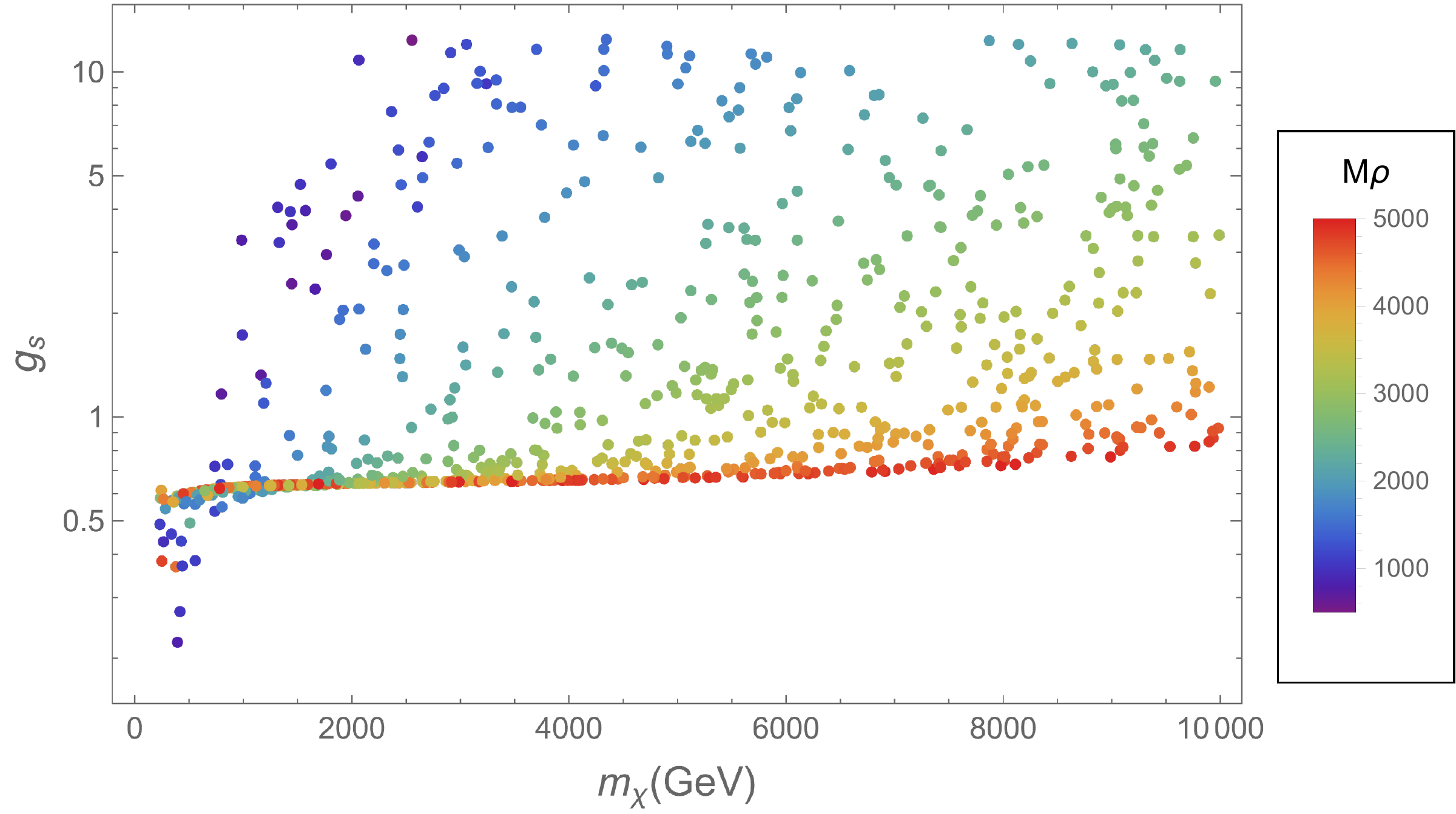}}
	\caption{The CDM coupling, $g_s$, versus the CDM mass. The quartic coupling constants is taken as 0.05. (a) $\sin\theta=0.01$  and (b) $\sin\theta=0.1$. }
	\label{fig1}
\end{figure}

A particle species in the early universe had interacted sufficiently with the other particles such that they were at thermal equilibrium with each other in the early cosmic soup. As the universe began to cool down, when its interaction rate dropped below the expansion rate of the universe, the equilibrium could no longer be maintained and the particle is said to be decoupled and its relic abundance remains constant until now. This scenario of thermal production of dark matter is known as the freeze-out mechanism. We assume that CDM particles were created through this mechanism.
The freeze-out temperature 
$T_{F}$
can be estimated through the iterative solution of the following equation \cite{kolb}:
\begin{equation}
x_{F}=\ln\bigg(\frac{m_{\chi}}{2 \pi^{3}}\sqrt{\frac{45 M_{Pl}}{2 g_{*}x_{F}}}\langle\sigma v \rangle\bigg),
\end{equation}
where
$ x_{F} \equiv\frac{m}{T_{F}} $,
$g_{*}$ 
is the effective degrees of freedom for the relativistic quantities in equilibrium and
$ M_{Pl} = 1.22\times 10^{19}$  GeV.  
Therefore accoding to this mechanism, the relic density  
$\Omega_{DM}h^{2}$
defined  as  the ratio of the present density of particles to the critical density,   can  be expressed as  follows: 
\begin{equation}
\Omega_{DM}h^{2}\approx\frac{(1.07\times10^{9})x_{F}}{\sqrt{g_{*}}M_{pl}(\mbox{GeV})\langle\sigma v \rangle},
\end{equation}
where 
  $\langle\sigma v \rangle $
   is the thermally averaged annihilation cross section times the relative velocity which is given by:
   \begin{equation}
 \langle\sigma v \rangle=\frac{1}{8 m_{\chi}^{4}T_{F}K_{2}^{2}(\frac{m_{\chi}}{T_{F}})}\int_{4m_{\chi}^{2}}^{\infty}\sigma_{\rm{ann}}(s)(s-4m_{\chi}^{2})\sqrt{s}K_{1}(\frac{\sqrt{s}}{T_{F}}) \,ds.
 \label{2}
   \end{equation}   
Here, $K_1(x)$ ($K_2(x)$) is the first (second) kind modified Bessel function.  We are interested in those regions of parameter space in which the CDM relic density satisfies the measured amount by the Plank observations (see Eq (1)). The other restrictions on the relevant parameter space come form the requirements of the vacuum stability and  perturbativity. The vacuum stability requirement imposes the following restrictions on the quartic coupling constants [22,23]:
	\begin{equation}
		\lambda>0,\hspace{1cm} 	\lambda_h>0,\hspace{1cm} 6\lambda_1+\sqrt{\lambda\lambda_h}>0.\label{sta}
	\end{equation}
	The perturbativity requirement leads to these upper restrictions [24]
	\begin{equation}
		\lambda<4\pi,\hspace{1cm} 	\lambda_h<16\pi,\hspace{1cm} \lambda_1<4\pi,\hspace{1cm}g_s<4\pi,\label{per}
	\end{equation} 
 where the last one reflects the perturbation requirement for the DM sector.
	We should note that the quartic coupling of SM Higgs is very loosely constrained by electroweak precision measurements. Therefore, new physics effects could induce large deviations from its SM expectation [25]. In order to explore the role of different parameters in this theory, we perform our analysis based on two different assumptions.  First, all parameters except for $g_s$, $m_\chi$ and $m_\rho$ are fixed. We chose $\sin\theta=0.01$ (panel a) and $\sin\theta=0.1$ (panel b) in Fig. \ref{fig1}. These values for $\theta$ are motivated because it has been
		shown that if the mass of singlet Higgs is about 750 GeV, $\sin\theta$ is constrained to be equal or less than 0.1 \cite{9}. 
		Second, the quartic Higgs coupling constants are allowed to vary in  regions consistent with extreme limits introduced in Eqs. (\ref{sta}) and (\ref{per}) in such a way that $m_\rho$ and $\theta$ are fixed (Fig. 2). Here, the gray points are related to $g_s>4\pi$ which do not obey the perturbativity constraint and we put them away for calculations. The red and blue points correspond to $\sin\theta=0.01$ and $\sin\theta=0.1$, respectively. Moreover, these regions must be consistent with the other observations including direct and indirect detection.
In the case of the former, it was shown that the leading order is far from the recent experimental sensitivities \cite{7}, however, the contribution of the leading loop correction can be surveyed by the upcoming experiments such as XENONnT \cite{77}. Now, we want to calculate the annihilation cross section of CDM particles into two monochromatic photons to check the consistency of related parameter space with the indirect searches such as H.E.S.S. experiment.

\begin{figure}[ht]
	\centering
	\subfigure[]{\includegraphics[scale=0.35]{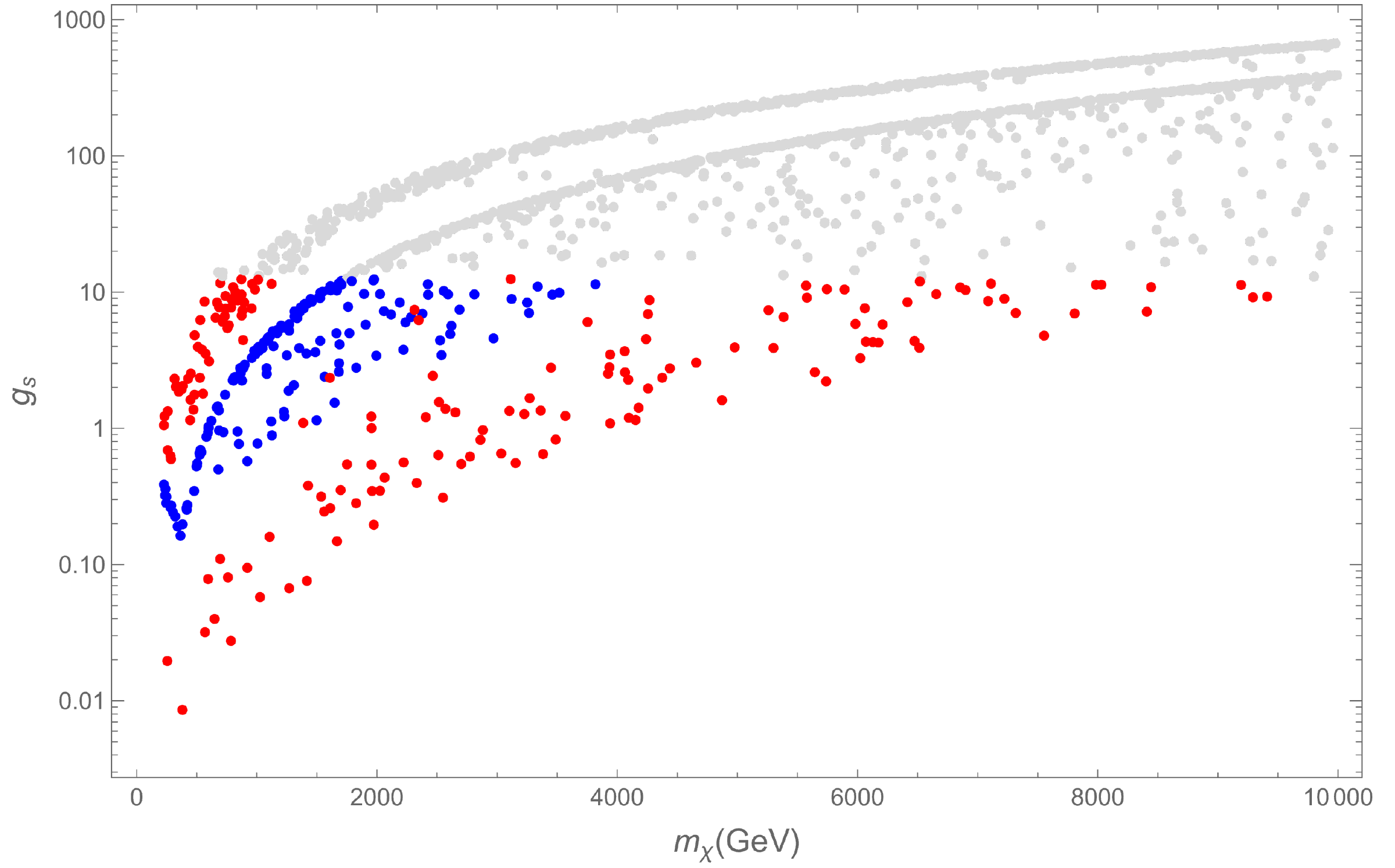}}\hspace{0.3cm}\subfigure[]{\includegraphics[scale=0.37]{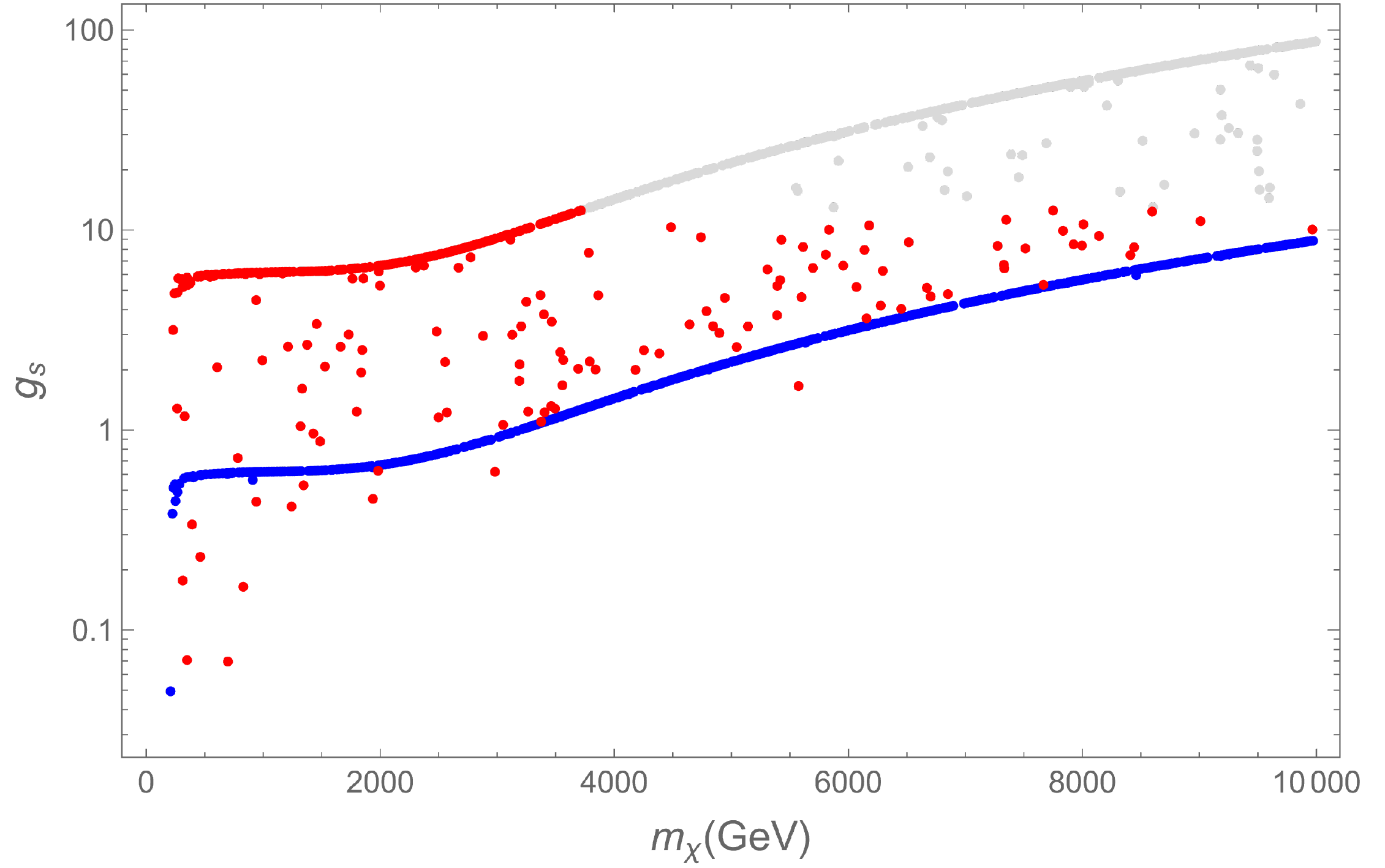}}
	\caption{The CDM coupling, $g_s$, versus the CDM mass. The quartic coupling constants are allowed to vary in the theoretically permissible region. Red and blue points correspond to  $\sin\theta=0.01$ and $\sin\theta=0.1$, respectively. The gray points are $g_s>4\pi$. (a) $m_\rho=750$ GeV and (b) $m_\rho=2.5$ TeV.}\label{fig222}
\end{figure}

\section{Annihilation  of CDM particles into a pair of photons} 
The H.E.S.S. experiment investigates the cosmic gamma-rays in the photon energy range of 0.03 to 100 TeV so it can observe high energy processes in the universe. In particular, it provides the ability to  tests actively unproven theories for dark matter through looking for predicted gamma-ray annihilation signals from CDM particles.
Since we have not practically detected any CDM particle until now, this technique as an ``indirect detection" of CDM is of interest. Indirect searches focus mainly on photons, neutrinos and antimatter cosmic-rays, notably positrons and antiprotons.
Among these, neutrinos and photons have advantages in comparison to others because they keep their source information during the streaming. Moreover, the very small cross sections of the neutrinos make their flux very difficult to detect. Therefore,  the   monochromatic lines of photon
are one of the promising messengers in the indirect search for CDM. 

Annihilation of CDM into the SM particles proceeds through $s$ channel with Higgs bosons exchange. The vertex of interaction between Higgs and pair photons comes from the quantum loop levels involving massive charged particles. The dominant Feynman diagrams contributing to this process are shown in Fig. \ref{fig2}. The cross section is written as follows:
	\begin{figure}[th]
	\includegraphics[width=9cm]{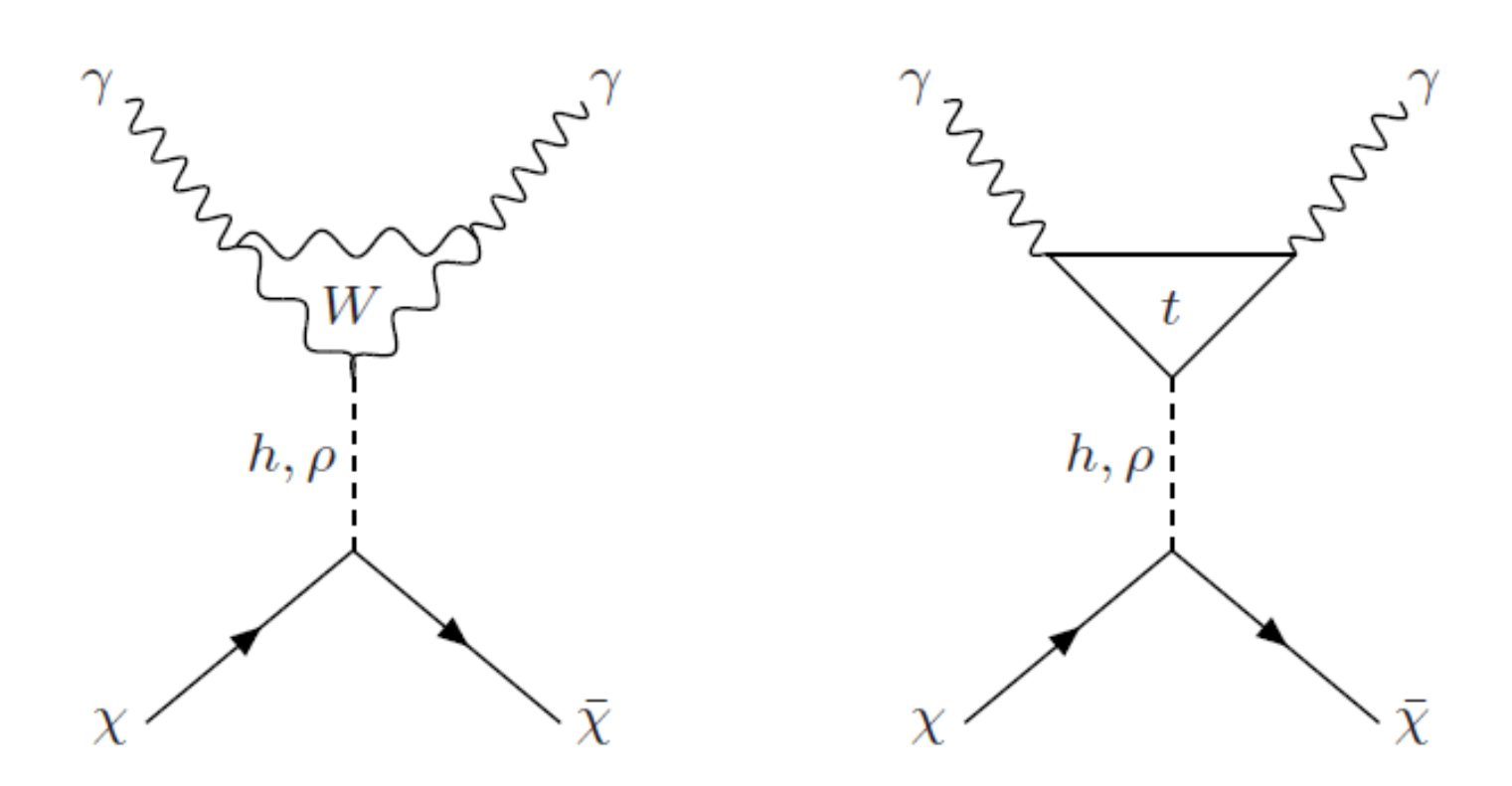}
	\caption{The dominant Feynman diagrams for the annihilation of a CDM pair into monochromatic gamma-ray lines.}
	\label{fig2}
\end{figure}

\begin{equation}
\sigma (s) v= \frac{1}{8\pi s} \frac{1}{4}\sum_{\rm{spins}}\lvert M_{\overline{\chi}\chi\to\gamma\gamma}\rvert^{2},
\end{equation} 
where $\sqrt{s}$
is the  center of mass energy. The amplitude $M_{\overline{\chi}\chi\to\gamma\gamma}$ is given by
 \begin{equation}
M_{\overline{\chi}\chi\to\gamma\gamma}(s)= \overline{v}(p)i g_s\gamma^5 u(p)\left[\frac{i \sin\theta}{s-m_{h}^{2}-i m_{h}\Gamma_{h}} M_{h\to\gamma\gamma}(s)+
\frac{i \cos\theta}{s-m_{\rho}^{2}-i m_{\rho}\Gamma_{\rho}} M_{\rho\to\gamma\gamma}(s)\right]
\end{equation}   
where $M_{{h,\rho}\to\gamma\gamma}$ are the decay amplitudes of Higgs particles to a pair of photons and one can write them as follows:
 \begin{eqnarray}
M_{h\to\gamma\gamma}&=&\frac{\alpha g_{e} s\cos\theta}{8\pi M_{W}}[3(\frac{2}{3})^{2}F_{t}+F_{W}]\\
M_{\rho\to\gamma\gamma}&=&\frac{\alpha g_{e} s\sin\theta}{8\pi M_{W}}[3(\frac{2}{3})^{2}F_{t}+F_{W}].
\end{eqnarray}
Here, $g_{e}=2\sqrt{\pi \alpha}/\sin\theta_{w}$  
and
 \begin{equation*}
 F_{t}= - 2 \tau [1+(1-\tau)f(\tau)] ,  
 \end{equation*}
  \begin{equation*}
  F_{W}= 2 + 3 \tau + 3\tau (2 -\tau) f(\tau),
  \end{equation*}
where
  $\tau=\frac{4 m_{i}^{2}}{s}$ 
   with 
   $i = t,W $
   and
  \begin{equation} 
  f(\tau)=\begin{cases}
  \left(\sin^{-1}\sqrt{\frac{1}{\tau}}\right)^{2} ,  & \mbox{if } \tau\geq1 \\ -\frac{1}{4}\left(\ln\frac{1+\sqrt{1-\tau}}{1-\sqrt{1+\tau}}-i\pi\right)^{2}, & \mbox{if }\tau<1  .
 \end{cases}
 	\end{equation}
Therefore, one can obtain the following expression for the annihilation cross section  of a pair CDM of particles into 
 	two photons:
 	\begin{equation}
 	\sigma (s)v= \frac{g_s^{2}}{32 \pi s}(2s)\bigg\lvert\\\frac{ \sin\theta M_{h\to\gamma\gamma}}{{s-m_{h}^{2}-i m_{h}\Gamma_{h}}}+\frac{ \cos\theta M_{\rho\to\gamma\gamma}}{{s-m_{\rho}^{2}-im_{\rho}\Gamma_{\rho}}}\bigg\rvert^{2}.
 	\end{equation}
Since dark matter is cold, we estimate $s \approx 4 m_{\chi} ^{2}$.	 We illustrate the cross section of annihilation into two photons versus the mass of CDM particle through Figs. 4 and 5, corresponding to the two different assumptions we used to extract $g_s$ in the previous section. 
	Furthermore, for comparison to the  H.E.S.S. data, we show this bound by a dashed line in these Figures. This study shows that this theory does not conflict with H.E.S.S. data. Indeed, the lower the mass of the second Higgs boson, the closer the theoretically predicted cross section to the experimental limit.

  	\begin{figure}[th]
  		\centering
  		\subfigure[]
  	{\includegraphics[scale=0.36]{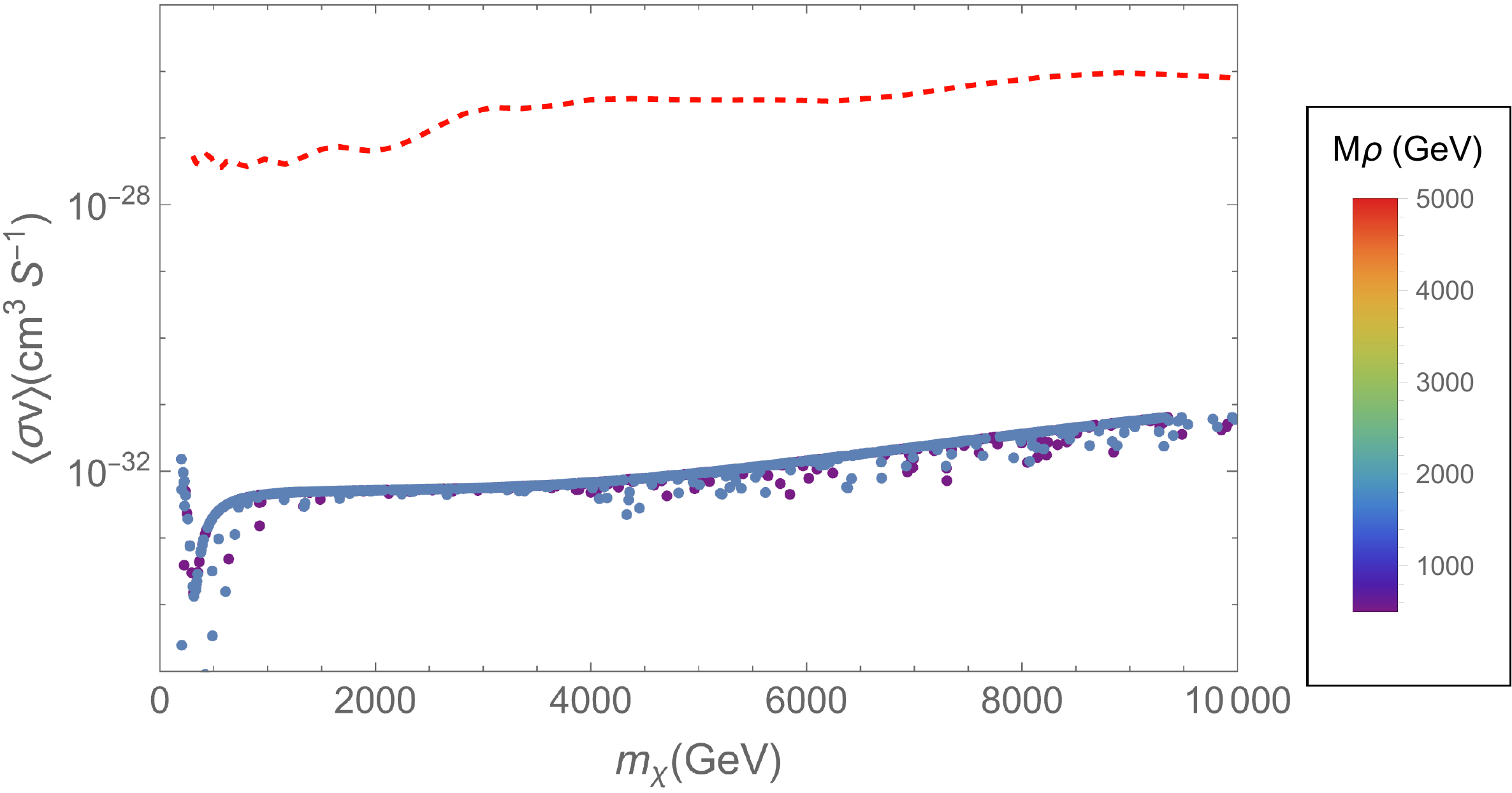}}\hspace{0.5cm}\subfigure[]{\includegraphics[scale=0.35]{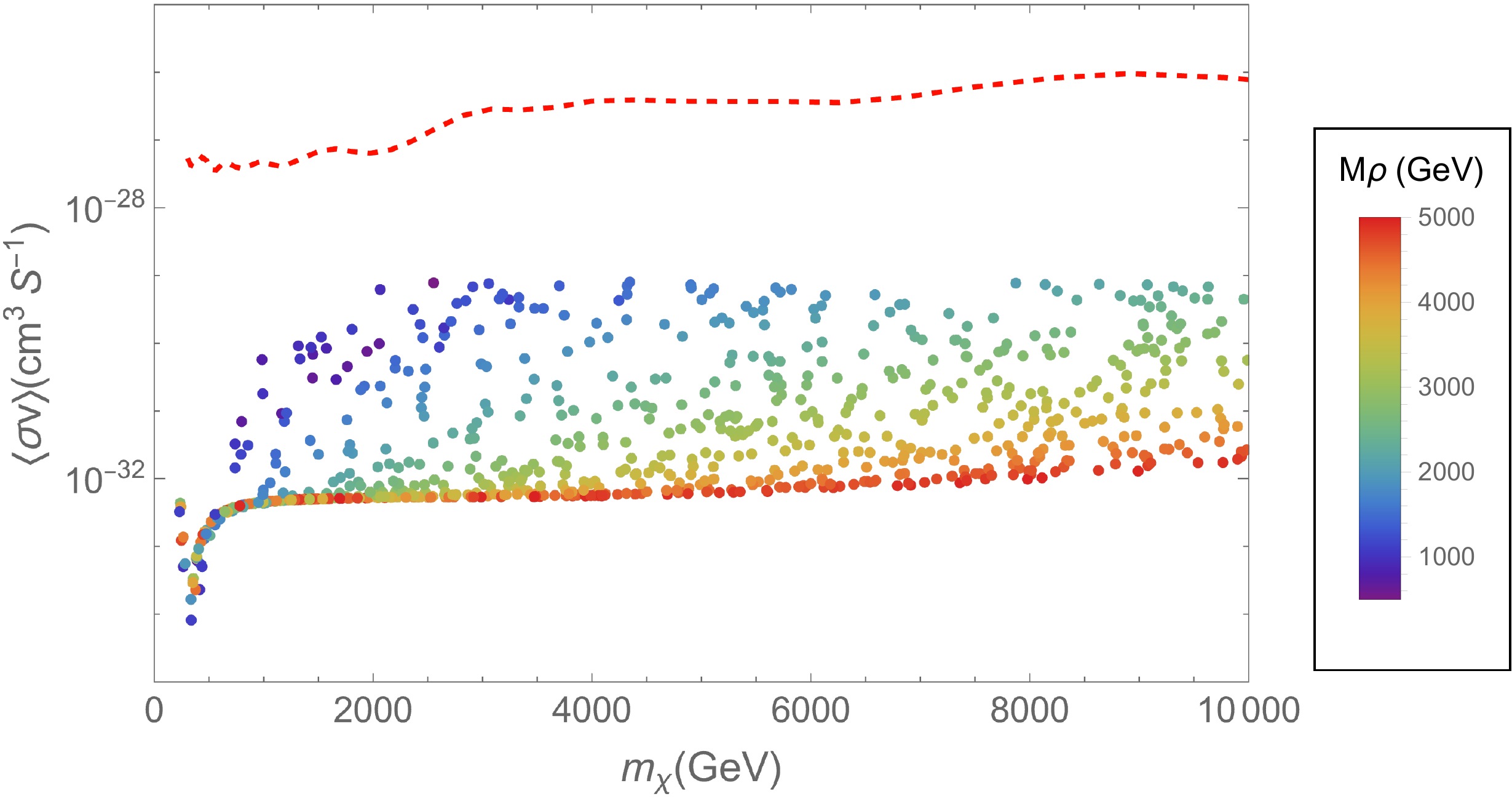}}
  	\caption{The monochromatic two photons annihilation of CDM particles in terms of the CDM mass and the second Higgs mediator $m_\rho$. The quartic coupling constants is taken as 0.05 and $g_s<4\pi$. The H.E.S.S. upper bound \cite{8} is showed by dashed line. (a) $\sin\theta=0.01$  and (b) $\sin\theta=0.1$.}
  	\label{fig3}
  \end{figure}
\begin{figure}[ht]
	\centering
	\subfigure[]{\includegraphics[scale=0.33]{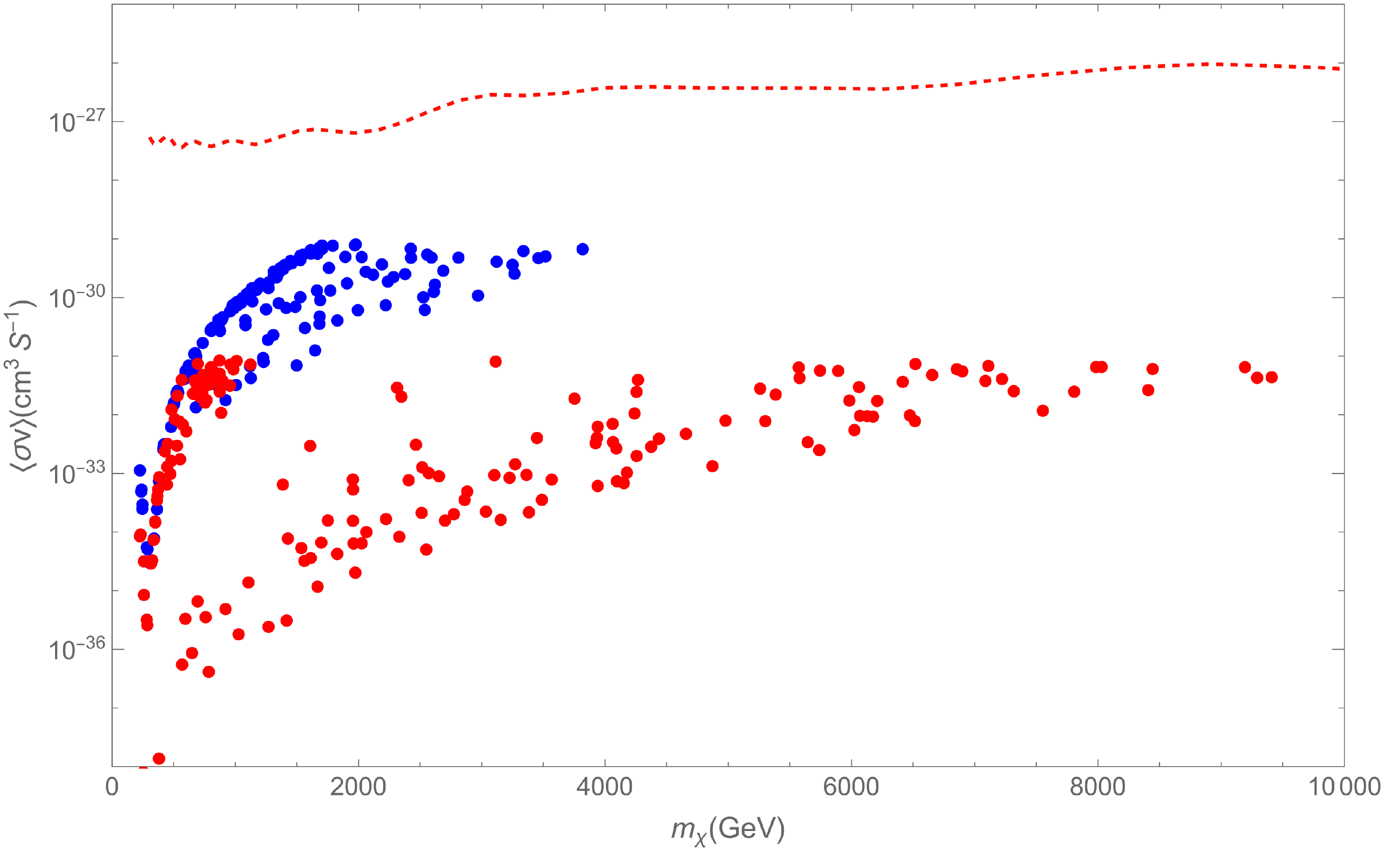}}\subfigure[]{\includegraphics[scale=0.36]{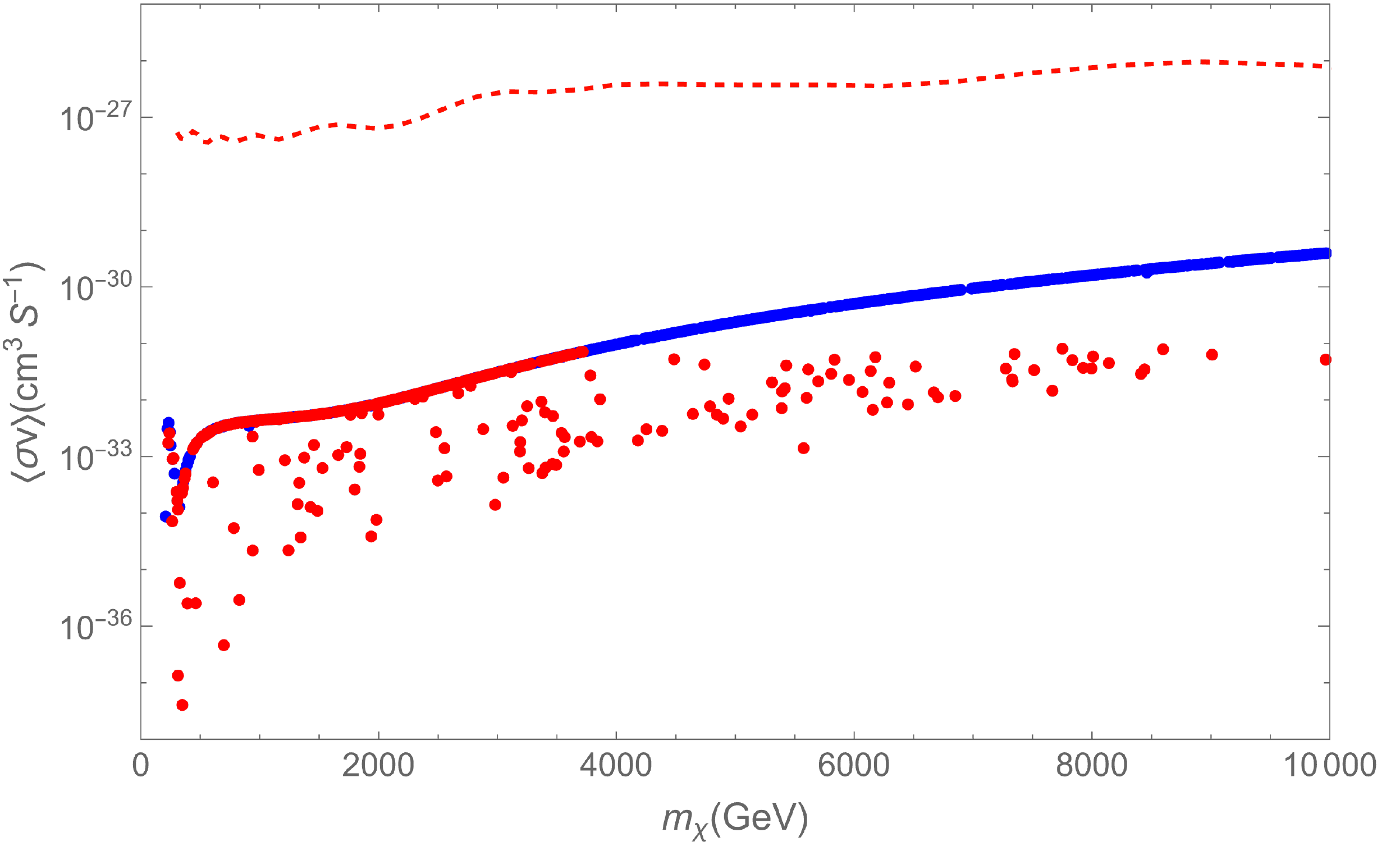}}
	\caption{The monochromatic two photons annihilation of CDM particles in terms of the CDM mass. The quartic Higgs coupling constants and $g_s$ are allowed to vary in the theoretically permissible region. Red and blue points correspond to  $\sin\theta=0.01$ and $\sin\theta=0.1$, respectively. (a) $m_\rho=750$ GeV and (b) $m_\rho=2.5$ TeV.}\label{fig333}
\end{figure}

\section{ Summary and conclusions}
We have considered an extension of SM involving a singlet fermion as cold dark matter candidate and a pseudo-scalar as mediator. In this theory, there exist some admissible parameter space regions which are consistent with the relic abundance and direct detection constraints \cite{7,Bastian,77}. Here, we have studied the monochromatic gamma-ray annihilation of dark matter particles within this theory.
The variation of the corresponding cross section in terms of the most relevant parameters has been illustrated in Figs. \ref{fig3} and \ref{fig333}.  Moreover, the experimental bounds on this cross section obtained by H.E.S.S experiment is depicted by dashed line in these figures. We see that the cross section of monochromatic gamma-ray annihilation does not conflict with the experimental bounds for all permissible parameter regions. Meanwhile, when the mass of the second Higgs boson is smaller than $1$ TeV by a greater distance, the theoretically predicted value for the cross section is closer to the experimental limit.

\vspace{5mm}

\appendix
\section{}
In this Appendix, we give the cross section of annihilation CDM into the SM particles and new Higgs. The cross section of SM particles except for Higgs is given by
\begin{eqnarray}
\sigma_{\rm{SM}}=&&\frac{(g_s\sin2\theta)^{2}}{64\pi  v_{\rm{rel}}}\left|\sum _{j=h,\rho} \frac{i}{s-m_j^2+{im}_j \Gamma _j}\right|^2
\nonumber\\&&\qquad\qquad\times\bigg[3\times 2s\left(\frac{m_{b}}{v_{H}}\right)^{2}\left(1-\frac{4m_{b}^{2}}{s}\right)^{\frac{3}{2}}+3\times 2s\left(\frac{m_{t}}{v_{H}}\right)^{2}\left(1-\frac{4m_{t}^{2}}{s}\right)^{\frac{3}{2}}+\left(\frac{m_{W}^{2}}{v_{H}}\right)^{2}\nonumber\\&&\qquad\qquad\qquad\left(2+\frac{(s-2m_{W}^{2})^{2}}{4m_{W}^{4}}\right)\left(1-\frac{4m_{W}^{2}}{s}\right)^{\frac{1}{2}}+\frac{1}{2}\left(\frac{m_{Z}^{2}}{v_{H}}\right)^{2}\left(2+\frac{(s-2m_{Z}^{2})^{2}}{4m_{Z}^{4}}\right)
	\left(1-\frac{4m_{Z}^{2}}{s}\right)^{\frac{1}{2}}\bigg],
\end{eqnarray}
where $v_{\rm{rel}}=\frac{1}{2\sqrt{1-4m_\chi^2/s}}$. The cross section of two Higgs bosons has three parts; into two SM Higgs bosons, into two new Higgs bosons and to different Higgs bosons:
\begin{equation}
	\sigma_{\rm{2Higgs}}=\frac 1 2 \sum_{i,j=1,2}\sigma_{h_ih_j},
\end{equation}
where $\sigma_{h_1h_2}$, in which the subscripts $h_1$ and $h_2$ stand for $h$ and $\rho$ in general, is given by
\begin{eqnarray}
	\sigma_{{h_1h_2}}&=&  \frac{{g_s}^2}{8 \pi s^2}\sqrt{E_1^2-m_1^2}\Bigg\{s\bigg|\sum _{j=h,\rho} \frac{{ig}_{{h_1h_2j}} S_j}{s-m_j^2+{im}_j \Gamma _j}\bigg|^2\notag\\
	&&\qquad+{g_s}^2 S_h^2 S_\rho^2\bigg\{  
	\sum _{j=h,\rho} \Bigg[\left(\frac{\epsilon_j \left(2 {EE}_j-m_j^2\right)  }{ {kp} \left[m_1^2+m_2^2-2 E \left(E_1+E_2\right)\right]}-\frac{1}{2 {kp}}\right)\left(2 {EE}_j-m_j^2+2 E^2\right)\ln A_j\notag\\&&\qquad\qquad
	-\frac{4 k^2 p^2}{\left(2 {EE}_j-m_j^2\right){}^2-4 k^2 p^2}+\frac{4 E\left[2 E \left({EE}_j-m_j^2\right)+{EE}_j^2-2 E_j k^2\right]}{\left(2 {EE}_j-m_j^2\right){}^2-4 k^2 p^2}\Bigg]\notag\\&&\quad
	\hspace{6cm}+\frac{2 \ln \left(A_1 A_2\right) k\left[2 {EE}_2-\left(p^2+2 E_1 E_2\right)\right]}{\left[m_1^2+m_2^2-2E \left(E_1+E_2\right)\right] p}-6\bigg\}\Bigg\}.
\end{eqnarray}
where $S_h=\sin\theta$ and $S_\rho=\cos\theta$, and $\epsilon_{\rho,h}=\pm1$. Here,  
$A_j=\frac{2EE_j-m_j^2+2kp}{2EE_j-m_j^2-2kp}$ where $E=\sqrt{s}/4$ and $k=\sqrt{E^2-m_{\chi}^2}$ are the energy and momentum of one of initial CDM particles in center of mass frame, and 
$E_h=\frac{4E^2+m_h^2-m_\rho^2}{4E}$, 
$E_\rho=\frac{4E^2+m_\rho^2-m_h^2}{4E}$ and $p=(16E^4-8m_h^2E^2-8m_\rho^2E^2-2m_h^2m_\rho^2+m_h^4+m_\rho^4)^{1/2}/4E$ are the energies and momentum of the final particles, respectively. The relevant vertex factors are given by
\begin{equation}
g_{hhh}=6 \left(\frac{1}{4} \lambda_H v_H \cos ^3\theta +\frac{1}{2} \lambda_1 \left(12 v_H \sin ^2\theta  \cos\theta +12v_\phi\sin\theta \cos^2\theta\right)+\lambda  v_\phi \sin ^3\theta\right),
\end{equation}
\begin{equation}
	g_{\rho\rho\rho}=6 \left(-\frac{1}{4} \lambda_H v_H  \sin^3\theta+\frac{1}{2} \lambda_1 \left(12 v_\phi \sin^2\theta\cos \theta-12 v_H \sin\theta\cos^2\theta\right)+\lambda   v_\phi \cos ^3\theta\right),
\end{equation}
\begin{equation}
	g_{hh\rho} = 
	2 \left(-\frac{1}{4} \lambda_H v_H \cos^2\theta \sin\theta + \lambda v_\phi \cos\theta \sin^2\theta+ 
	\frac{1}{2} \lambda_1 \left(4 v_\phi \cos^3\theta + 
	8 v_H \cos^2\theta\sin\theta- 
	8 v_\phi\cos\theta\sin^2\theta - 
	4 v_H\sin^3\theta\right)\right),
\end{equation}
and
\begin{eqnarray}
	g_{\rho \rho h}=2 \Bigg(&&\hspace{-0.4cm}\frac{1}{4} \lambda_H v_H \sin ^2(\theta ) \cos\theta+\lambda_1 \cos\theta \left(2 v_H \cos ^2\theta-4 v_\phi \sin \theta \cos\theta\right)+\lambda_1 \sin\theta \left(2 v_\phi \sin ^2\theta-4 v_H \sin\theta\cos\theta\right)\nonumber\\
	&&+\lambda  v_\phi \sin\theta \cos ^2\theta\Bigg),
\end{eqnarray}
in which
\begin{equation}
	v_\phi=\frac{1}{2 \lambda }\left(-6 \lambda_1 v_H \cot\theta+6 \lambda_1 v_H \tan \theta+\cot\theta\sqrt{24 \lambda\lambda_H v_H^2 \tan ^2\theta+\lambda_1^2 v_H^2 \left(6-6 \tan ^2\theta \right)^2} \right).
\end{equation}

\end{document}